\documentclass[conference]{IEEEtran}
\IEEEoverridecommandlockouts
\usepackage{cite}
\usepackage{amsmath,amssymb,amsfonts}
\usepackage{graphicx}
\usepackage{textcomp}
\usepackage{xcolor}
\usepackage{optidef}
\usepackage{graphicx}
\usepackage{balance}
\usepackage{float}
\usepackage{algorithm}
\usepackage{algpseudocode}
\def\BibTeX{{\rm B\kern-.05em{\sc i\kern-.025em b}\kern-.08em
    T\kern-.1667em\lower.7ex\hbox{E}\kern-.125emX}}
\begin{document}

\title{Harnessing the Potential of Omnidirectional UAVs in RIS-Enabled Wireless Networks
\thanks{This document has been produced with the financial assistance of the European Union (Grant no. DCI-PANAF/2020/420-028), through the African Research Initiative for Scientific Excellence (ARISE), pilot programme. ARISE is implemented by the African Academy of Sciences with support from the European Commission and the African Union Commission. The contents of this document are the sole responsibility of the author(s) and can under no circumstances be regarded as reflecting the position of the European Union, the African Academy of Sciences, and the African Union Commission.}
}

% \and
% \and

\author{
  \IEEEauthorblockN{
    Abdoul Karim A. H. Saliah$^1$,
    Hajar El Hammouti$^1$,
    Daniel Bonilla Licea$^{1,2}$
  }

    \IEEEauthorblockA{
   $^1$College of Computing, Mohammed VI Polytechnic University (UM6P), Benguerir, Morocco.
  }

     \IEEEauthorblockA{
    $^2$Faculty of Electrical Engineering, Czech Technical University in Prague, Czech Republic.
  }

 \IEEEauthorblockA{
      Email: \{abdoul.saliah, hajar.elhammouti, daniel.bonilla\}@um6p.ma}
}

\maketitle

\begin{abstract}
Multirotor Aerial Vehicles (MRAVs) when integrated into wireless communication systems and equipped with a Reflective Intelligent Surface (RIS) enhance coverage and enable connectivity in obstructed areas. However, due to limited degrees of freedom (DoF), traditional under-actuated MRAVs with RIS are unable to control independently both the RIS orientation and their location, which significantly limits network performance. A new design, omnidirectional MRAV (o-MRAV), is introduced to address this issue. In this paper, an o-MRAV is deployed to assist a terrestrial base station in providing connectivity to obstructed users. Our objective is to maximize the minimum data rate among users by optimizing the o-MRAV's orientation, location, and RIS phase shift. To solve this challenging problem, we first smooth the objective function and then apply the Parallel Successive Convex Approximation (PSCA) technique to find efficient solutions. Our simulation results show significant improvements of {\bf 28\%} and {\bf 14\%} in terms of minimum and average data rates, respectively, for the o-MRAVs compared to traditional u-MRAVs.
\end{abstract}

\begin{IEEEkeywords}
Communication-aware robotics, UAVs, omnidirectional MRAV (o-MRAV), reflective Intelligent Surface (RIS), relays, UAVs. 
\end{IEEEkeywords}

%%%%%%%%%%%%%%%%%%%%%%%%%%%%%%%%%%%%%%%%%%%%%%%%%%%%%%%%%%%%%%%%%%%%%%%%%%%%%%%%%%%%%%%%%%%%%%%%%%%%%%%%%%%%%%%%%%%%%
%-I.Introduction
%%%%%%%%%%%%%%%%%%%%%%%%%%%%%%%%%%%%%%%%%%%%%%%%%%%%%%%%%%%%%%%%%%%%%%%%%%%%%%%%%%%%%%%%%%%%%%%%%%%%%%%%%%%%%%%%%%%%%
\section{Introduction}

In the past decade, the incorporation of unmanned aerial vehicles (UAVs) into wireless networks has attracted substantial interest from both academia and industry\cite{Saliah2024MSAG}. UAVs have been explored for a variety of applications, including improving network coverage, disaster recovery, and data collection~\cite{BonillaPIEEE2024, MozaffariIEEECST2019, ndiaye2022age,Ndiaye2023DataFreshness}. Within this context, a substantial portion of research has focused on multirotor aerial vehicles (MRAVs), with particular emphasis on a design known as under-actuated MRAVs (u-MRAVs)~\cite{HamandiIJRR2021}. These UAVs are characterized by having fewer actuation inputs—such as thrust, pitch, roll, and yaw angles—than degrees of freedom (DoF), which include their 3D position and orientation. As a result,  the 3D orientation of u-MRAVs depends on their translational velocity, making it impossible to achieve independent control of both 3D position and 3D orientation~\cite{HamandiIJRR2021,Bonilla2024OMRAV_Mag}. 

%, including serving as rapid replacements for damaged base stations in post-disaster scenarios, acting as efficient relays to collect data from Internet of Things (IoT) devices, and functioning as aerial base stations to enhance network capacity in densely populated or underserved areas 

 Recently, the robotics community has introduced new MRAV designs, referred to as omnidirectional MRAVs (o-MRAVs). These designs allow the total thrust vector to be controlled in all directions, independently of torque, with multiple input combinations. This results in more actuation inputs than degrees of freedom, which enables complete and independent control over each movement axis, and allows precise orientation management \cite{HamandiIJRR2021,Bonilla2024Globecom,BonillaICASSP2024}.

To further enhance the versatility of UAVs in wireless networks, another promising technology has been considered: reconfigurable intelligent surfaces (RIS). RIS can be deployed to efficiently redirect signals toward users in obstructed areas and improve overall communication performance. This has given rise to the concept of aerial RIS (ARIS), which aims to dynamically position RIS in 3D space using a UAV. %By equipping an o-MRAV with a RIS, the o-MRAV can mechanically control the orientation of the RIS and its position to ensure an optimal signal redirection towards obstructed users.

Several works have investigated the 3D position of an ARIS for multiple purposes. For instance, RIS placement and 3D passive beamforming are optimized in~\cite{lu2021aerial} to enhance the worst signal-to-noise ratio. In~\cite{jeon2022energy}, a high aerial platform (HAP) enabled by RIS is examined for redirecting backhaul signals to UAVs. The authors propose a joint optimization of positioning and RIS phase shifts to improve system energy efficiency. Additionally, the case of multiple ARIS is explored in~\cite{aung2023energy}, where deep reinforcement learning is applied to jointly optimize ARIS placement, phase shifts, and power control. 
However, only a limited number of research papers have explored the impact of RIS orientation. In~\cite{cheng2022ris}, the rotation of the RIS plane is examined to enhance network capacity. Similarly, in~\cite{zeng2020reconfigurable}, the authors investigate the optimal orientation of a RIS to improve the data rate for a single user. Additionally,~\cite{elzanaty2021reconfigurable} focuses on adjusting RIS orientation to improve localization accuracy in indoor environments. %Notably, all the studies have employed RIS mounted on fixed platforms or u-MRAVs.

%\cite{zeng2020reconfigurable} (communication letter with one user)

%\cite{cheng2022ris} (rotation) 

%\cite{elzanaty2021reconfigurable} (localization)

This paper, motivated by the previous studies on the effect of the orientation of the RIS, explores the deployment of o-MRAVs equipped with RIS and shows the benefits that the optimization of the RIS orientation and position brings to the network performance. In particular, we analyze how control over the orientation of the RIS, in addition to phase shift and 3D position, can significantly enhance the minimum data rate for users in obstructed environments. To the best of our knowledge, this is the first work to explore the integration of RIS with o-MRAVs and to offer new insights into their potential to enhance wireless networks.

%\clearpage

\textit{Notations}: $y$ is a scalar, $\boldsymbol{y}$ or $\boldsymbol{Y}$ is a vector or matrix. $\boldsymbol{Y}^T$ and $\boldsymbol{Y}^*$ denote the transpose, and Hermitian of $\boldsymbol{Y}$, respectively.  $|.|$ denotes the modulus of a complex number and $\|.\|_2$ denotes the $l_2$-norm. $j$ is the imaginary unit. 
%$\theta_{_{W-Z}}$ and $\xi_{_{W-Z}}$ denote an elevation angle and an azimuth angle of the channel between a transmitter $W$ and a receiver $Z$, respectively.
\( \text{mod}(.,.) \) is the modulus and \( \left\lfloor . \right\rfloor \) is the floor function.

%%%%%%%%%%%%%%%%%%%%%%%%%%%%%%%%%%%%%%%%%%%%%%%%%%%%%%%%%%%%%%%%%%%%%%%%%%%%%%%%%%%%%%%%%%%%%%%%%%%%%%%%%%%%%%%%%%%%%
%-II.System Model
%%%%%%%%%%%%%%%%%%%%%%%%%%%%%%%%%%%%%%%%%%%%%%%%%%%%%%%%%%%%%%%%%%%%%%%%%%%%%%%%%%%%%%%%%%%%%%%%%%%%%%%%%%%%%%%%%%%%%
%\vspace{-0.66cm}
\section{System Model}

\begin{figure}[h]
    \centering
    \includegraphics[width=0.93\linewidth]{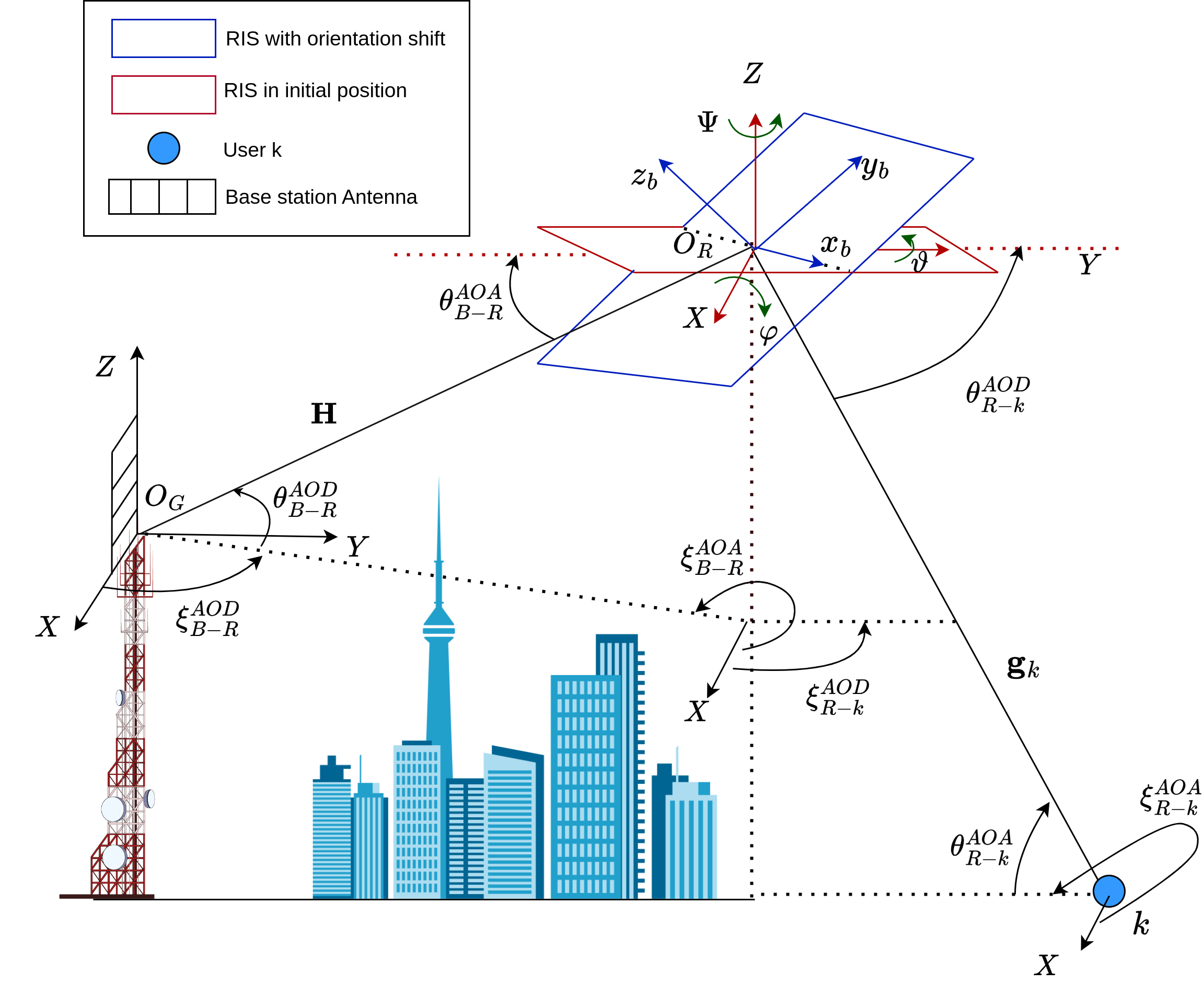}
    \caption{Communication Model} 
    \label{fig:Channel_Archi}
\end{figure}

We consider a communication model where a base station~$B$ communicates with $K$ user equipments (UEs) in an obstructed environment. As a result, there exists no direct link between the base station and the UEs, as illustrated in Fig.\ref{fig:Channel_Archi}. To overcome this problem, an o-MRAV equipped with a RIS ensures the scattering of the signal received from the base station towards the UEs. We also assume that the base station is equipped with a linear antenna array separated by a vertical distance $d_{V_0}$, and the set of antenna elements is denoted by $\mathcal{N}=\{1,\ldots,N\}$. To reflect the signal, the o-MRAV carries an RIS consisting of $M = M_H \times M_V$ elements, arranged in a uniform rectangular array. The horizontal and vertical separation distances between RIS elements are denoted by $d_{H_R}$ and $d_{V_R}$, respectively. The set of RIS elements is given by $\mathcal{M} = \{1,\ldots,M\}$ and the set of UEs is denoted by $\mathcal{K}=\{1,\ldots,K\}$. For the sake of notation, in the remainder of the paper, we designated $k$ to represent a user, $R$ to represent the RIS, and $B$ to represent the base station.

We consider two reference coordinate systems: the global reference system $(O_G, X, Y, Z)$, centered at the base station antenna, and the mobile body reference system $(O_R,x_b, y_b, z_b)$, centered at the MRAV center. In the global reference system, the positions of the base station antenna, the MRAV, and any user equipment (UE) $k \in \mathcal{K}$ are denoted by $\mathbf{p}_B = [x_B, y_B, z_B]^T$, $\mathbf{p}_R = [x_R, y_R, z_R]^T$, and $\mathbf{p}_k = [x_k, y_k, z_k]^T$, respectively. Additionally, the position of an individual antenna element $n \in \mathcal{N}$ in the global reference system is expressed as
\vspace{-0.1cm}
\begin{equation}
    \mathbf{p}_n^B=\left[d_{H_0},0,(n-1) d_{V_0}\right]^T, \forall n \in \mathcal{N},
\end{equation}
where $d_{H_0}$ represents the horizontal width of the BS antenna array. Similarly, the position of a RIS element $m \in \mathcal{M}$ in the mobile body reference system is given by~\cite{Sherman2023AoI}
\vspace{-0.09cm}
\begin{equation}
    \mathbf{p}_m^R\!=\!\left[\lfloor \!(m-1\!)/M_V \rfloor d_{V_R}\!,\mathrm{mod}\!\left(\!(m-1)\!,M_H\!\right)d_{H_R}\!,0\right]^T\!.
\end{equation}

%The positions $\mathbf{p}_B=[x_B,y_B,z_B]^T$, $\mathbf{p}_R=[x_R,y_R,z_R]^T$, and $\mathbf{p}_k=[x_k,y_k,z_k]^T$ are the positions of the base station antenna, the MRAV, and a UE $k \in \mathcal{K}$ in the global reference system, respectively. The position $\mathbf{p}_n^B$ of an antenna element in the global reference system is expressed as follows
%\begin{equation}
%    \mathbf{p}_n^B=\left[d_{H_0},0,(n-1) \times d_{V_0}\right]^T, \forall n \in \mathcal{N}.
%\end{equation}

%The position $\mathbf{p}_m^R$ of a RIS element $m$ in the body reference system is expressed as in \cite{Sherman2023AoI} by
%\begin{equation}
%    \mathbf{p}_m^R = \begin{bmatrix}
%    \lfloor (m-1)/M_V \rfloor d_{V_R} \\
%    mod\left((m-1),M_H\right)d_{H_R} \\
%    0
%    \end{bmatrix}.
%\end{equation}

The o-MRAV is capable of changing its orientation independently of its position. To describe its orientation, we use Euler angles with $XYZ$ convention: roll $\left(\varphi\right)$, pitch $\left(\vartheta\right)$, and yaw $\left(\psi\right)$. Consequently, the orientation of the o-MRAV is represented as a vector $\boldsymbol{\Omega}=\left[\varphi, \vartheta, \psi\right]^{\top} \in \mathbb{R}^3$. 

The orientation $\boldsymbol{\Omega}$ directly impacts the positions of RIS elements in the global reference system. Specifically, the vector $\mathbf{p}_{1-m}^R(\boldsymbol{\Omega})$, which represents the position of any RIS element $m$ relative to the first RIS element in the global reference system is given by
%\vspace{-0.23cm}
\begin{equation}
    \mathbf{p}_{1-m}^R(\boldsymbol{\Omega}) = \mathcal{R}_{\boldsymbol{\Omega}}(\mathbf{p}_m^R - \mathbf{p}_1^R), \forall m \in \mathcal{M},
\end{equation}

\noindent with $\mathcal{R}_{\boldsymbol{\Omega}}$ the rotation matrix from the body reference to the global reference, which is given by
\begin{equation}
\mathcal{R}_{\boldsymbol{\Omega}} = \left[\begin{array}{ccc}
c_{\vartheta} c_{\psi} & -c_{\vartheta} s_{\psi} & s_{\vartheta} \\
c_{\varphi} s_{\psi} + c_{\psi} s_{\varphi} s_{\vartheta} & c_{\varphi} c_{\psi} - s_{\varphi} s_{\vartheta} s_{\psi} & -c_{\vartheta} s_{\varphi} \\
s_{\varphi} s_{\psi} - c_{\varphi} c_{\psi} s_{\vartheta} & c_{\psi} s_{\varphi} + c_{\varphi} s_{\vartheta} s_{\psi} & c_{\varphi} c_{\vartheta}
\end{array}\right],
\normalsize
\end{equation}
\noindent where $s_{.}$ and $c_{.}$ are respectively sine and cosine functions.

We define the channel between the base station and the o-MRAV as $\mathbf{H} \in \mathbb{C}^{N \times M}$, and the channel between the o-MRAV and any given user $k$ as $\mathbf{g}_k \in \mathbb{C}^{M \times 1}$. They are expressed as
\vspace{-0.09cm}
\begin{equation}
    \mathbf{H} = \eta_{_{B-R}}\mathbf{a}_{rx,R}\left(\theta_{B-R}^{AOA}, \xi_{B-R}^{AOA}, \mathbf{\Omega} \right) \mathbf{a}_{tx,B}^*\left(\theta_{B-R}^{AOD}, \xi_{B-R}^{AOD}\right),
\end{equation}
\vspace{-0.15cm}
\begin{equation}
    \mathbf{g}_k = \eta_{_{R-k}} \mathbf{a}_{tx,R}^*\left(\theta_{R-k}^{AOD}, \xi_{R-k}^{AOD}, \mathbf{\Omega}\right),
\end{equation}
\noindent with $\eta_{_{B-R}}=\frac{\sqrt{\beta_0}}{d_{B-R}} e^{\frac{-j 2 \pi}{\lambda} d_{B-R}}$ and $\eta_{_{R-k}}=\frac{\sqrt{\beta_0}}{d_{R-k}} e^{\frac{-j 2 \pi}{\lambda} d_{R-k}},$

\noindent where $\beta_0$ is the reference path gain at a distance of 1 meter. The distances $d_{B-R} = \|\mathbf{p}_R - \mathbf{p}_k\|_2$ and $d_{R-k} = \|\mathbf{p}_k - \mathbf{p}_R\|_2$ represent the distances between the base station and the RIS, and between the RIS and user $k$, respectively. The angles $(\theta_{B-R}^{AOD}, \xi_{B-R}^{AOD})$ and $(\theta_{B-R}^{AOA}, \xi_{B-R}^{AOA})$ denote the angles of departure and arrival for the channel between the base station and the RIS. Similarly, $(\theta_{R-k}^{AOD}, \xi_{R-k}^{AOD})$ represent the angles of departure for the channel between the RIS and user $k$. $\mathbf{a}_{tx,B}$, $\mathbf{a}_{rx,R}$ and $\mathbf{a}_{tx,R}$ denotes respectively the transmit array response of the base station, RIS receive array response and RIS transmit array response. They are given by 
\vspace{-0.15cm}
\begin{equation}
\begin{split}
    \mathbf{a}_{tx,B}\left(\theta_{B-R}^{AOD}, \xi_{B-R}^{AOD}\right) = \Big[ & e^{j \mathbf{s}(\theta_{B-R}^{AOD}, \xi_{B-R}^{AOD}) \mathbf{p}_{1}^B}, \\
    & \ldots, e^{j \mathbf{s}(\theta_{B-R}^{AOD}, \xi_{B-R}^{AOD}) \mathbf{p}_{N}^B} \Big]^T,
\end{split}
\end{equation}

\begin{equation}
\begin{split}
    \mathbf{a}_{rx,R}\left(\theta_{B-R}^{AoA}, \xi_{B-R}^{AoA}, \mathbf{\Omega}\right) = \Big[ & e^{j \mathbf{s}(\theta_{B-R}^{AoA}, \xi_{B-R}^{AoA}) \mathbf{p}_{1-1}^R(\boldsymbol{\Omega})}, \\
    & \ldots, e^{j \mathbf{s}(\theta_{B-R}^{AoA}, \xi_{B-R}^{AoA}) \mathbf{p}_{1-M}^R(\boldsymbol{\Omega})} \Big]^T,
\end{split}
\end{equation}
\vspace{-0.2cm}
\begin{equation}
\begin{split}
    \mathbf{a}_{tx,R}\left(\theta_{R-k}^{AoD}, \xi_{R-k}^{AoD}, \mathbf{\Omega}\right) = \Big[ & e^{j \mathbf{s}(\theta_{R-k}^{AoD}, \xi_{R-k}^{AoD}) \mathbf{p}_{1-1}^R(\boldsymbol{\Omega})}, \\
    & \ldots, e^{j \mathbf{s}(\theta_{R-k}^{AoD}, \xi_{R-k}^{AoD}) \mathbf{p}_{1-M}^R(\boldsymbol{\Omega})} \Big]^T,
\end{split}
\end{equation}

\noindent where $    \mathbf{s}(\theta_{_{W-Z}}, \xi_{_{W-Z}})$ represents a wave vector, which indicates the phase variation of a plane wave at a given position in the global reference. Its expression is given by\cite{Bjornson2017} as follows
\vspace{-0.09cm}
\begin{equation}
\begin{split}
    \mathbf{s}(\theta_{_{W-Z}}, \xi_{_{W-Z}}) = \frac{2 \pi}{\lambda} \Big[ & \cos (\theta_{_{W-Z}}) \cos (\xi_{_{W-Z}}), \\
    & \cos (\theta_{_{W-Z}}) \sin (\xi_{_{W-Z}}), \; \sin (\theta_{_{W-Z}}) \!\Big],
\end{split}
\end{equation}
where $\theta_{_{W-Z}}$ and $\xi_{_{W-Z}}$ denote the elevation angle and the azimuth angle, respectively, of the channel between the transmitter $W$ and the receiver $Z$.

Accordingly, the data rate of a given user $k$ is defined by
\begin{equation}
    R_k = B_k\log_2 \left(1 + \frac{P_0 G_A\left|\mathbf{g}_k\mathbf{M}_{\mathbf{\Theta}}\mathbf{H} \mathbf{f}\right|^2}{ \sigma_0^2} \right) ,\forall k \in \mathcal{K},
\end{equation}

\noindent where $B_k$ is the bandwidth allocated to user $k$, $P_0$ is the power of transmission of the base station, $G_A$ the gain of the base station antenna, the reflection matrix of the RIS is $\mathbf{M}_{\mathbf{\Theta}} = \text{diag}(\{e^{j\theta_i}\}_{i=1}^M)$ with $\mathbf{\Theta}=(\theta_1,\ldots,\theta_M)$, $\mathbf{f}$ is the beamforming vector at the base station and $\sigma_0^2$ is the variance of an additive white Gaussian
noise. We consider that a maximum ratio strategy (MRT) is applied to maximize the signal-to-noise ratio at the base station\cite{Zheng2022IRS}\cite{jeon2022energy}. This results in beamforming that depends only on the array response at the base station through the following expression \cite{jeon2022energy}
\begin{equation}
\mathbf{f}=\mathbf{a}_{tx,B}\left(\theta_{B-R}^{AOD}, \xi_{B-R}^{AOD}\right)/{\|\mathbf{a}_{tx,B}\left(\theta_{B-R}^{AOD}, \xi_{B-R}^{AOD}\right)\|_2}.
\end{equation}

\noindent Furthermore, we assume, for the sake of tractability, that the channels $\mathbf{H}$ and $\mathbf{g_k}$ are known at the base station. This can be achieved through the use of channel estimation techniques \cite{Wei2021Estimation}. Additionally, we consider the use of frequency division multiple access (FDMA) at the base station to ensure that there is no interference.

%%%%%%%%%%%%%%%%%%%%%%%%%%%%%%%%%%%%%%%%%%%%%%%%%%%%%%%%%%%%%%%%%%%%%%%%%%%%%%%%%%%%%%%%%%%%%%%%%%%%%%%%%%%%%%%%%%%%%
%-III.Minimum Rate Optimization
%%%%%%%%%%%%%%%%%%%%%%%%%%%%%%%%%%%%%%%%%%%%%%%%%%%%%%%%%%%%%%%%%%%%%%%%%%%%%%%%%%%%%%%%%%%%%%%%%%%%%%%%%%%%%%%%%%%%%
\vspace{-0.2cm}
\section{Minimum Rate Optimization}\label{Optimization}

This work aims to maximize the minimum rate among users by optimizing the o-MRAV orientation and position in addition to the RIS reflection matrix. The purpose of this work is to demonstrate that incorporating o-MRAV orientation alongside position and phase shift optimization yields significant gains compared to traditional approaches with fixed orientation. The optimization problem is expressed as follows
\vspace{-0.25cm}
\begin{maxi!}|s|
{\mathbf{\Omega}, \mathbf{p_R}, \mathbf{\Theta}}{\left(\min_{k \in \mathcal{K}} R_k\right)}{}{}
\tag{\theequation}
\addConstraint{\varphi \in \left[0, \frac{\pi}{2}\right], \vartheta \in \left[0, \frac{\pi}{2}\right], \psi \in \left[0, 2\pi\right) \label{Const1}}{}{}
\addConstraint{\theta_m \in [0, 2\pi], \forall m \in \mathcal{M} \label{Const2}}{}{}
\addConstraint{\mathbf{p_R}^T \in \left[x_{\text{min}}, x_{\text{max}}\right] \times \left[y_{\text{min}}, y_{\text{max}}\right] \times \left[z_{\text{min}}, z_{\text{max}}\right] .\label{Const3}}{}{}
\end{maxi!}

In this problem, constraint (\ref{Const1}) limits the UAV orientation to keep the RIS visible to the base station and users, (\ref{Const2}) defines the range of the phase shift of RIS elements and constraint (\ref{Const3}) delimits the region covered by the UAV.

This problem is a challenging one as it is a non-convex one. Moreover, the minimum term in the maximization problem further complicates the problem because of the non-smooth nature of the min function. 

To overcome this problem, we first approximate the min function using the $p$-norm, which provides a close approximation when $p\rightarrow -\infty$ \cite{Boyd2004}.  The studied problem is then reformulated as
\vspace{-0.22cm}
\begin{maxi!}|s|
{\mathbf{\Omega}, \mathbf{p_R}, \mathbf{\Theta}}{\left\|\left(R_1, \ldots, R_K\right)\right\|_p}{}{}
\tag{\theequation}
\label{objFunc}
\addConstraint{(\ref{Const1}),(\ref{Const2}),(\ref{Const3})}{}{}
\end{maxi!}

We consider the convex feasible set of solutions of the problem to be $\mathcal{X} = \mathcal{X}_1 \times \mathcal{X}_2 \times \mathcal{X}_3 $,  where $\mathcal{X}_1 = [0, 2\pi]^M$, $\mathcal{X}_2 = \left[x_{\text{min}}, x_{\text{max}}\right] \times \left[y_{\text{min}}, y_{\text{max}}\right] \times \left[z_{\text{min}}, z_{\text{max}}\right]$, and $\mathcal{X}_3 = [0, \frac{\pi}{2}]^2 \times [0, 2\pi]$ are respectively the RIS phase shift, the MRAV position and orientation. We then obtain the following problem 

\vspace{-0.5cm}

\begin{mini!}|s|
{\mathbf{x}_1, \mathbf{x}_2, \mathbf{x}_3}{F(\mathbf{x_1}, \mathbf{x_2}, \mathbf{x_3}) = -\left\|\left(R_1, \ldots, R_K\right)\right\|_p}{}{}
\tag{\theequation}
\label{objFunc_PSCA}
\addConstraint{\mathbf{x}_i \in \mathcal{X}_i, \forall i \in \{1,2,3\},}{}{}
\end{mini!}

% Parallel SCA
\begin{algorithm}[t]
\small
\caption{Parallel Successive Convex Approximation (PSCA)}
\begin{algorithmic}[1]
\State Set $l = 0$, initialize with a feasible point $\mathbf{x}^0 \in \mathcal{X}$, and $\{\gamma^l\} \in (0, 1]$.
\Repeat
    \ForAll{$i \in \{1,2,3\}$ \textbf{(in parallel)}}
        \State Solve (\ref{SurrogateFunc})
    \EndFor
    \State Compute the next iterate:
    \State $\mathbf{x}^{l+1} = \mathbf{x}^l + \gamma^l (\hat{\mathbf{x}}(\mathbf{x}^l) - \mathbf{x}^l)$
    \label{Step7}
    \State $l \gets l + 1$
\Until{convergence}
\State \textbf{return} $\mathbf{x}^l$
\end{algorithmic}
\label{AlgoPSCA}
\end{algorithm}

\noindent This results in a non-convex multi-block optimization problem, which we address using the Parallel Successive Convex Approximation (PSCA) method.

The PSCA as described in Algorithm \ref{AlgoPSCA} consists of iteratively solving convex approximations of the objective function for all blocks. First, at each iteration $l$, the following convex problems are solved in parallel 

\begin{equation}  \hat{\mathbf{x}}_i(\mathbf{x}^l) = \arg \min\limits_{\mathbf{x}_i \in \mathcal{X}_i} \tilde{F}_i(\mathbf{x}_i \mid \mathbf{x}^l), \forall i \in \{1,2,3\}
    \label{SurrogateFunc}
\end{equation}

\noindent where $\tilde{F}_i(\mathbf{x}_i \mid \mathbf{x}^l)$ is the first-order Taylor expansion of $F$ around $\mathbf{x}^l$ with respect to $\mathbf{x}_i$. Second, the next iterate $\mathbf{x}^{l+1}$ is computed in step \ref{Step7} by linear combination of the obtained optimal $\hat{\mathbf{x}}_i(\mathbf{x}^l)$and $\mathbf{x}^l$. This process is repeated until convergence. The complexity analysis of the PSCA method is thoroughly discussed in \cite{Razaviyayn2014PSCA}.

%SCA
% \begin{algorithm}[t]
% \caption{SCA (Successive Convex Approximation)}\label{Algo}
% \small
% \begin{algorithmic}[1]
% \State Set $l = 0$, initialize with  $\mathbf{\Theta}^0 \in \mathcal{X}$, $\{\gamma^l\} \in (0, 1]$
% \Repeat
%     \State Solve surrogate problem: 
%       (\ref{FTE_F})
%     \State Compute next iterate : 
%    $$\mathbf{\Theta}^{l+1} = \mathbf{\Theta}^l + \gamma^l \left( (\mathbf{\Theta}_{opt}(\mathbf{\Theta}^{l}) - \mathbf{\Theta} \right)$$ \label{step4}\vspace{-0.55cm}
%     \State $l \gets l + 1$
% \Until{convergence}
% \State \Return $\mathbf{\Theta}^l$
% \end{algorithmic}
% \end{algorithm}

%figures
\begin{figure*}[htbp]
\centering
\begin{tabular}{ccc}
  \includegraphics[scale=0.5]{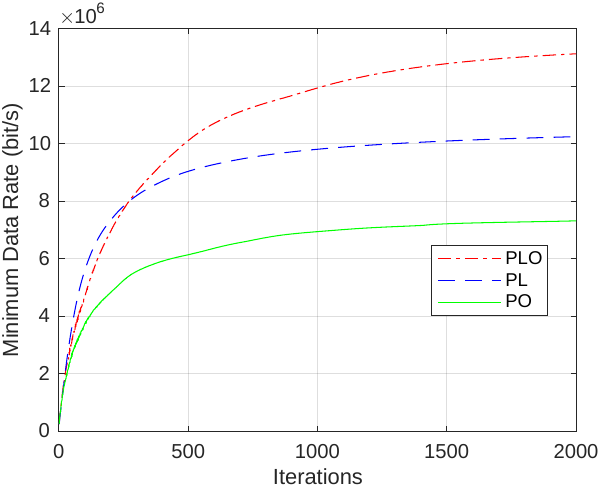} & 
  \includegraphics[scale=0.5]{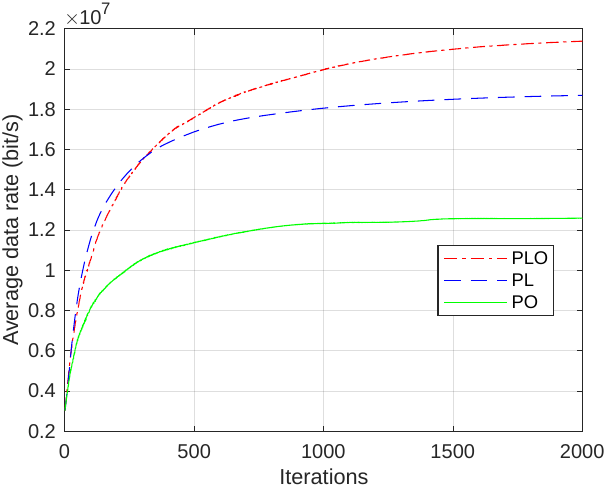} & 
  \includegraphics[scale=0.5] {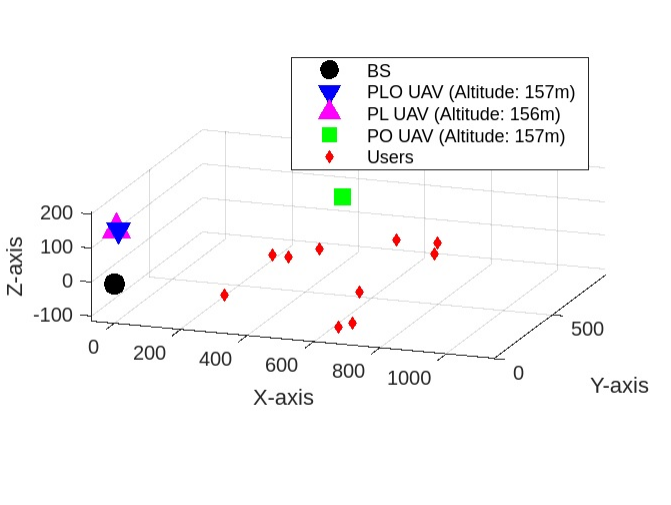} \\
  \small (a) & 
  \small (b) & 
  \small (c) 
\end{tabular}
\caption{(a) Minimum rate vs iterations, (b) Average rate vs iterations, (c) Optimized positions of o-MRAV. }
\label{fig:Rates_Position}
\end{figure*}

\begin{figure}[htbp]
    \centering
    \includegraphics[trim={0 0.5cm 2cm 1cm},clip,width=0.8\linewidth,height=0.4\linewidth]{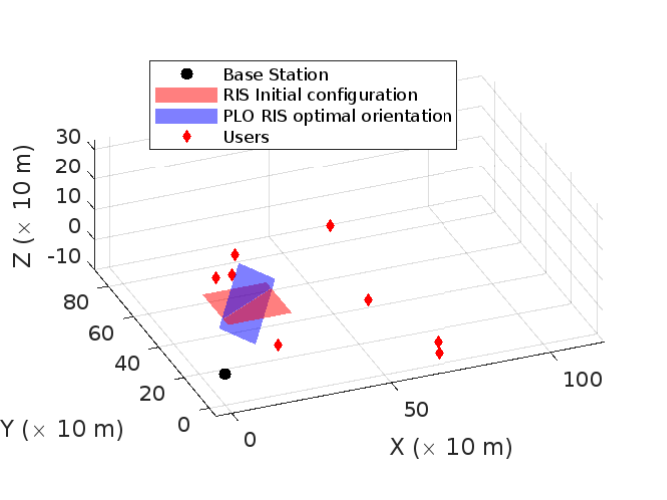}
    \caption{Optimal RIS orientation for PLO.}
    \label{fig:Orientation}
\end{figure}

%%%%%%%%%%%%%%%%%%%%%%%%%%%%%%%%%%%%%%%%%%%%%%%%%%%%%%%%%%%%%%%%%%%%%%%%%%%%%%%%%%%%%%%%%%%%%%%%%%%%%%%%%%%%%%%%%%%%%
%-IV.Simulation Results
%%%%%%%%%%%%%%%%%%%%%%%%%%%%%%%%%%%%%%%%%%%%%%%%%%%%%%%%%%%%%%%%%%%%%%%%%%%%%%%%%%%%%%%%%%%%%%%%%%%%%%%%%%%%%%%%%%%%%

\section{Simulation Results}

To assess the performance of an o-MRAV mounted with RIS, we consider an area of $1000 \text{m} \times 1000 \text{m}$ where $K=10$ users are randomly distributed on the ground. A base station at an altitude $h_{BS}=68m$ is positioned at the center of the global reference and has $N=10$ antenna elements.
The signal emitted by the base station has a wavelength of $\lambda=0.15 \text{m}$. The base station transmits with a power of $P_0=1 \text{W}$ and an antenna gain of $G_A=8 \text{dB}$.
The horizontal and vertical distances between the base station's antenna elements are $d_{H_0}=\frac{\lambda}{2}$ and $d_{V_0}=\frac{\lambda}{2}$, respectively. The RIS number of elements is $M_H\times M_V=5\times 4$, and the horizontal and vertical separation distances of the RIS are $d_{H_R}=\frac{\lambda}{2}$ and $d_{V_R}=\frac{\lambda}{2}$.
The o-MRAV coordinates \(x_R\), \(y_R\), and \(z_R\) are constrained to \(x_R \in [0, 1000m]\), \(y_R \in [0, 1000m]\), and \(z_R \in [150m, 300m]\), respectively.
 We suppose the reference gain at $1\text{m}$ is $\beta_0=-30 \text{dB}$, the noise power is $\sigma^2=-100 \text{dBm}$ \cite{HU2023Secure}, and the bandwidth of each user is $B_k=10 \text{MHz}$. For the $p$-norm  approximation in problem (\ref{objFunc_PSCA}), we consider $p=-8$.

%To assess the performance of the proposed approach, we considered the following benchmarks: 
We evaluate the following three approaches: i) PLO: This scheme is the proposed approach described in the section. \ref{Optimization}, where the RIS phase shift, The MRAV location, and orientation are all optimized by PSCA. ii) PL: This scheme consists in phase shift and location optimization using PSCA. It corresponds to traditional optimization found in the literature for UAV-carried RIS scenarios\cite{Hammouti2024EE}\cite{Zhai2022EE_RIS}. iii) PO: This approach consists of phase shift and RIS orientation optimization through PSCA, with a fixed position of MRAV set on top of the barycenter of all users, at the optimal altitude obtained through PLO. 
%\begin{itemize} 
%\item PLO: This scheme is the proposed approach described in the section. \ref{Optimization}, where the RIS phase shift, The MRAV location, and orientation are all optimized by PSCA.
%\item PLO-2: This corresponds to phase shift, location, and orientation optimization through the Matlab function \textit{fmincon}. 
%\item PL: This scheme consists in phase shift and location optimization using PSCA. It corresponds to traditional optimization found in the literature for UAV-carried RIS scenarios\cite{Hammouti2024EE}\cite{Zhai2022EE_RIS}. 
%\item PO: This approach consists of phase shift and RIS orientation optimization through PSCA, with a fixed position of MRAV set on top of the barycenter of all users, at the optimal altitude obtained through PLO. 
% \item PO-2: Similar to PO-1, this approach also optimizes the phase shift of the RIS and its orientation via  PSCA, with the position of the MRAV set on top of the barycenter of all users' positions but at an altitude corresponding to PL optimal UAV altitude. 
%\end{itemize}

In Fig.~\ref{fig:Rates_Position}(a), we plot the minimum data rate against the number of iterations, obtained by averaging over $10$ simulations. As can be seen, PLO achieved the best performance, followed by PL, and PO schemes. PLO achieved a minimum data rate of $28\%$, and $ 80\%$ high compared to PL and PO, respectively. %The under-performance of PLO-2 compared to PLO-1 indicates that $fmincon$ is less robust compared to PSCA. 
The reason of PL's under-performance compared to PLO is that in addition to phase shift and location, PLO also optimizes the orientation of the RIS, thanks to the o-MRAV's ability to control both position and orientation independently. The PO scheme's poor performance is due to its fixed MRAV location, which highlights the importance of optimizing the MRAV location first; orientation optimization further enhances performance when the location is optimal.

Fig.~\ref{fig:Rates_Position}(b) plots the average data rate of users against the number of iterations, obtain by averaging over $10$ simulations. The same trend as in Fig.~\ref{fig:Rates_Position}(a) is observed. PLO performs $14\%$ and  $69\%$ higher than PL and PO, respectively. This shows the overall gain achieved in terms of data rate through the joint optimization of MRAV location and RIS orientation. Additionally, by maximizing the minimum data rate among users, the network's average data rate is improved.

Fig.~\ref{fig:Rates_Position}(c) illustrates one of the network configurations under consideration, which shows the final positions achieved by the ARIS for all schemes. We observe that the MRAV positions for PLO and PL are near the base station at an altitude close to the minimum of $150\,\text{m}$ due to minimal path loss experienced by the received signal in this area. Thanks to the RIS, the signal can then be transmitted through passive beamforming to users over long distances. %In contrast, for the PO scheme, the position remains constant.

In Fig.~\ref{fig:Orientation}, the final orientation of the PLO RIS, for one of the network configurations under consideration, is plotted. The figure shows the scheme's ability to vary the RIS's orientation while maintaining visibility with both the base station and the users.

%%%%%%%%%%%%%%%%%%%%%%%%%%%%%%%%%%%%%% %%%%%%%%%%%%%%%%%%%%%%%%%%%%%%%%%%%%%%%%%%%%%%%%%%%%%%%%%%%%%%%%%%%%%%%%%%%%%%%
%-VI.Conclusion
%%%%%%%%%%%%%%%%%%%%%%%%%%%%%%%%%%%%%%%%%%%%%%%%%%%%%%%%%%%%%%%%%%%%%%%%%%%%%%%%%%%%%%%%%%%%%%%%%%%%%%%%%%%%%%%%%%%%%

%\vspace{-0.6cm}
\section{Conclusion}
%\vspace{-0.2cm}
In this paper, we investigate the use of o-MRAV deployed as relays for terrestrial communication in obstructed environments. The emphasis was on the joint optimization of the o-MRAV orientation and position, and the RIS phase shift to maximize the minimum data rate among users. To tackle the challenging studied problem, a close smooth approximation was used and then the problem was solved using PSCA. Through simulation experiments, we showed that o-MRAV independent orientation and location control enhance both the minimum rate and the average rate by $28\%$ and $14\%$, respectively, compared to traditional u-MRAV. In future works, we are planning to investigate o-MRAV-equipped RIS deployment to combat eavesdropping. We will also investigate the energy efficiency of o-MRAV-equipped RIS when used as relays to assist terrestrial communication.

\clearpage
\balance
\bibliographystyle{IEEEbib}
\bibliography{references}
\end{document}